# BIBLIOMETRIC ANALYSIS OF THE WORLD SCIENTIFIC PRODUCTION IN CHEMICAL ENGINEERING DURING 2000-2011. PART 3: ANALYSIS OF RESEARCH TRENDS AND HOT TOPICS

Ruben Miranda (✉)[1], Esther García-Carpintero[2]

[1]Departamento de Ingeniería Química y de Materiales, Facultad de CC. Químicas, Universidad Complutense de Madrid. Avda. Complutense s/n, 28040 Madrid-Spain rmiranda@ucm.es.
[2]Agencia de Evaluación de Tecnologías Sanitarias, Instituto de Salud Carlos III.
Avda. Monforte de Lemos 5, 28029 Madrid-Spain eegarcia@isciii.es

(✉)Corresponding author: rmiranda@ucm.es

## ABSTRACT

A comprehensive bibliometric analysis of the scientific production of Chemical Engineering area has been carried out using the Web of Science database for the period 2000-2011 through three complementary studies. Part 3 has analyzed the distribution of words in article titles, keyword plus® and author keywords of both total scientific production and the 1,000 most cited publications. The main areas of Chemical Engineering have been identified; they are mainly related to chemical reaction engineering such as catalysis, reactors, kinetics, and unit operations such as adsorption. Furthermore, a total of ten hotspots in the area have been identified: hydrogen as a new energy vector, wastewater treatments, carbon dioxide (capture and sequestration), photocatalysis, nanoparticles, biodiesel, nanotubes, ionic liquids, advanced oxidation processes, membranes, fuel cells and the use of biomass as raw material (e.g. bioethanol, energy production, etc.). Results obtained suggest thematic areas and research trends can be easier analyzed through the most cited publications, decreasing significantly the time necessary for these analyses. Words in article title, keyword plus® and author keywords are complementary, however, author keywords are suggested as the source of most useful data. Compared to other strategies, the author keywords are more valuable for identifying research areas and trends, and the total number of words to be analyzed is lower. Whatever the case, authors recommend a revision of the results obtained by experts in the area to avoid inaccurate results and get the most meaningful information.

**ÖZET**

Kimya Mühendisliği alanının bilimsel üretiminin kapsamlı bir bibliyometrik analizi, tamamlayıcı üç çalışma aracılığıyla 2000-2011 dönemi için Web of Science veritabanı kullanılarak gerçekleştirilmiştir. Bölüm 3, hem toplam bilimsel üretim hem de en çok alıntı yapılan 1.000 yayının makale başlıkları, keyword plus® ve yazar anahtar sözcüklerindeki kelimelerin dağılımını analiz etmektedir. Kimya Mühendisliğinin ana alanları belirlenmiştir; bunlar temel olarak kataliz, reaktörler, kinetik gibi kimyasal reaksiyon mühendisliği ve adsorpsiyon gibi birim işlemlerle ilgilidir. Ayrıca, bölgede toplam on yoğunlaşılan nokta tespit edilmiştir: yeni bir enerji vektörü olarak hidrojen, atık su arıtmaları, karbondioksit (yakalama ve ayırma), fotokataliz, nanopartiküller, biyodizel, nanotüpler, iyonik sıvılar, gelişmiş oksidasyon süreçleri, membranlar, yakıt hücreler ve biyokütlenin hammadde olarak kullanılması (örneğin biyoetanol, enerji üretimi, vb.). Elde edilen sonuçlar, tematik alanların ve araştırma eğilimlerinin en çok alıntı yapılan yayınlar aracılığıyla daha kolay analiz edilebileceğini ve bu analizler için gerekli süreyi önemli ölçüde azalttığını göstermektedir. Makale başlığındaki sözcükler, anahtar sözcük plus® ve yazar anahtar sözcükleri tamamlayıcıdır, ancak en yararlı verilerin kaynağı olarak yazar anahtar sözcükleri önerilmektedir. Diğer stratejilerle karşılaştırıldığında, yazar anahtar kelimeleri araştırma alanlarını ve eğilimlerini belirlemek için daha değerlidir ve analiz edilecek toplam kelime sayısı daha düşüktür. Durum ne olursa olsun, yazarlar, yanlış sonuçlardan kaçınmak ve en anlamlı bilgileri almak için alandaki uzmanlar tarafından elde edilen sonuçların revizyonunu önermektedir.

## 1. INTRODUCTION

Bibliometry is a research method used to evaluate research performance by the analysis of publication patterns, offering quantitative and objective information on the activity in science and technology. One application of bibliometry is to determine the previous, current, and future research trends or focus of a discipline or topic (Huang, 2009).

Research emphasis and trends can be analyzed from the words in the titles and abstracts or author keywords from scientific literature. The title of an article includes words which the authors consider as the most importance essence from their research work (Tanaka & Ho, 2011). The author keywords offer the information of research trends as viewed by researchers (Xie *et al.*, 2008; Li *et al.*, 2009). Furthermore, keywords plus® are index terms created by Thomson Reuters from significant, frequently occurring words in the titles of an article’s cited references, but do not necessarily appear in the title of the article itself (Zhang *et al.*, 2010a; Web of Science, 2014).

The analysis of words in article title and abstract, author keywords and keywords plus® were used to evaluate research trends in different areas and periods in the recent years (Xie *et al.,* 2008; Li *et al.,* 2009, Zhang *et al.,* 2010a, 2010b; Yi & Jie, 2011; Fu *et al.*, 2013, 2014; Tanaka & Ho, 2011; Wang *et al.,* 2011). Generally, in order to overcome the weak points of the three separate types of keywords analysis, the analysis of the words in article title, author keywords, and keywords plus® were combined. In this way, synonymic single words and congeneric phrases are summed into categories, so as to analyze the historical development of the science more completely and precisely and to discover the directions the science is taking (Tanaka & Ho, 2011).

Most of the previous bibliometric studies including research emphasis and trends analyses discussed briefly this topic and most of them are focused on total scientific production, however, to the best knowledge of the authors, they have never applied simultaneously to the most cited papers of the area. As the volume of the most cited papers is considerably lower than the total scientific production, usually 100-1,000 most cited papers (Ioannidis, 2006; Madhan *et al.*, 2010) or 1% total scientific production (which corresponds to the definition of ESI indicators from Web of Science) (King, 2004; Miyairi & Chang, 2012; Chuang *et al.*, 2013), it would be very interesting to analyze if comparable results can be obtained. This would considerably reduce the time necessary to carry out the research trends analyses, which is probably their main difficulty for these analysis, allowing their extended use.

Therefore, the objective of the present study is to comparatively analyze both the best source of documents to be used (total scientific production or only the most cited papers) and the most useful keywords (words in article title, author keyword or keyword plus® from Web of Science) for research emphasis and trends analysis in Chemical Engineering.

Chemical Engineering synthesizes knowledge form several disciplines and interacts with researchers from multiple disciplines. However, Chemical Engineering has demonstrated a unique ability to synthesize diverse forms of knowledge from applied sciences and other engineering disciplines into cohesive and effective solutions to many societal needs (National Research Council, 2007). In this sense, Chemical Engineering covers a wide-ranging set of social interests and needs, including health, habitable environment, transportation, communications, agriculture, clothing and food, national defense and security, and various life amenities (National Research Council, 2007). A 2.2% of total publications (articles and reviews) covered in SCI-Expanded in 2015 were classified in “Chemical Engineering” area, being the 14th subject category of SCI-Expanded in terms of publications.

Despite the scientific importance of Chemical Engineering, most of the previous bibliometric studies focused only in a specific geographical region or country (Yin, 2009; Fu *et al.*, 2014; García-Carpintero & R. Miranda, 2013; Rojas-Sola & San-Antonio-Gómez, 2010a, 2010b; Chang & Cheng, 2012), specific journals (Schubert, 1998) or only in the highly cited papers in the area (Ho, 2012; Chuang *et al.*, 2013). Furthermore, only some of these studies included brief analyses of the thematic areas and research trends in the area.

## 2. METHODOLOGY

Parts 1 and 2 of this study described the methodology followed to obtain the bibliographic records considered for total scientific production and the 1,000 most cited papers in Chemical Engineering area (Miranda and Garcia-Carpintero, 2020a, 2020b). Briefly, the study has been circumscribed to the journals indexed in the area of “Engineering, Chemical” of the Web of Science, the number of journals covered in this area varying from 110 to 135 during the analyzed period (2000-2011). It is evident that although the journals indexed in the thematic area of “Engineering, Chemical” do not represent 100% of the scientific production in Chemical Engineering, they represent a good sample of the research carried out in this area (Rojas-Sola and San Antonio Gómez, 2010a, 2010b) (see Part 1 of this study for further details). Only two document types were considered: articles and reviews.

The bibliographic information of all the documents, 214,264 for total scientific production and 1,000 for the highly cited papers, were downloaded from the Web of Science. The words in title, the author keywords and keyword plus® were analyzed both for total scientific production and the 1,000 cited papers separately. All the bibliographic records were further processed in Microsoft Excel® before analysis. First, some format corrections were carried out such as: the removal of spaces before the keywords, the transformation of all the keywords to capital letters, among others. Next, Microsoft Excel® tools were used for automatic counting of the different words year by year. However, further manual work were always necessary, i.e. the combination of singular and plurals, including common abbreviations in the counting, the consideration of chemical formulas and common names for chemical substances, etc.

The analysis of keywords, especially those of author keywords, was very challenging. The keywords of an article are not single words but a combination of several words. This makes difficult the automatic counting of the more used keywords. In addition, there are several combinations of words which means the same, i.e. “membrane filtration”, “membrane”, etc. Furthermore, when using compounds, several ways to say the same can be found, i.e. “methane” or “CH4”, “hydrogen” and “H2”, “titanium dioxide”, “TiO2” or “titania”, etc. There are also some words with different spelling depending on British and American English, i.e. “modeling” (US) and “modelling” (UK), “aluminum” (US) and “aluminium” (UK), etc. Some of these words were those with the highest frequency in publications. Furthermore, another difficulty presented was that around 2/3 parts of the words appeared only once. This makes necessary a thorough analysis of the results given as the initial number of counts for a term, can be increased even 10 times after the revision of the results. In this sense, synonyms must be taken into account to have the most representative values of the research area of the articles. In many cases, therefore is recommended the obtained results to be reviewed by experts in the field.

## 3. RESULTS AND DISCUSSION

### 3.1. Analysis of total publications

Words in article title. There were analyzed 214,264 articles with title (100% of total publications). The average number of single words in the title of the articles was 12.0, increasing slightly with time, from 10.9 in 2000 to 12.8 in 2011. The total number of words analyzed was 2,558,032, most of them appearing only once (61.5%) appeared only once. An important amount of words were prepositions, conjunctions, etc. such as “of”, “and”, “the”, “in”, “a”, “for”, etc. (representing 28.1% of the total number of raw words counted). These words together with other useless words to determine research trends such as “effect”, “study” or “analysis”, were also removed before analysis.

Table 1 shows the 30 most frequently single words used in article titles. The first two words in the rank were “membrane” and “catalyst”, which indicates that the use of membranes for separation processes and the use of catalysts for chemical reactions are two of the most important areas in Chemical Engineering. These areas shown an increase in importance during the analyzed period. In 2000, “membrane” and “catalyst” appeared in 5.0% and 4.3% of the articles analyzed, while these values increased to 6.4% and 5.9%, respectively, in 2011.

**Table 1.** Top 30 most frequently words in the title, keyword plus® and author keywords in Chemical Engineering area during the period 2000-2011. Note: singular and plurals of the words were counted together. NP = number of publications; % TP= percentage of total publications.

| Rank | **Words in title** | **NP (% TP)** | **Keyword plus®** | **NP (% TP)** | **Author keyword** | **NP (% TP)** |
|---|---|---|---|---|---|---|
| 1 | "Membrane" | 13665 (6.4%) | "System" | 9640 (5.7%) | "Modeling" / "Modelling" | 3691 (2.6%) |
| 2 | "Catalyst" | 12528 (5.9%) | "Model" | 8435 (5.0%) | "Adsorption" | 3686 (2.6%) |
| 3 | "Process" | 12221 (5.7%) | "Oxidation" | 8359 (5.0%) | "Kinetics" | 2981 (2.1%) |
| 4 | "System" | 11314 (5.3%) | "Water" | 7991 (4.7%) | "Optimization" / "Optimisation" | 1871 (1.3%) |
| 5 | "Property" | 9442 (4.4%) | "Kinetics" | 7369 (4.4%) | "Simulation" | 1825 (1.3%) |
| 6 | "Model" | 9148 (4.3%) | "Adsorption" | 6598 (3.9%) | "CFD" / "computational fluid dynamics" | 1812 (1.3%) |
| 7 | "Modeling" | 8989 (4.2%) | "Performance" | 5688 (3.4%) | "Mass transfer" | 1731 (1.2%) |
| 8 | "Acid" | 8860 (4.2%) | "Temperature" | 5687 (3.4%) | "Carbon dioxide" / "CO2" | 1634 (1.1%) |
| 9 | "Synthesis" | 8753 (4.1%) | "Removal" | 5684 (3.4%) | "Photocatalysis" / "Photocatalytic" | 1536 (1.1%) |
| 10 | "Water" | 8642 (4.1%) | "Behavior" | 5404 (3.2%) | "Ultrafiltration" / "UF" | 1520 (1.1%) |
| 11 | "Carbon" | 8112 (3.8%) | "Carbon-dioxide" / "CO2" | 5292 (3.1%) | "Membrane" | 1433 (1.0%) |
| 12 | "Gas" | 8031 (3.8%) | "Mixtures" | 5177 (3.1%) | "Mathematical modeling/modelling" | 1281 (0.90%) |
| 13 | "Solution" | 7397 (3.5%) | "Separation" | 5079 (3.0%) | "Nanofiltration" / "NF" | 1151 (0.81%) |
| 14 | "Kinetic" | 7268 (3.4%) | "Flow" | 5053 (3.0%) | "Reverse osmosis" / "RO" | 1110 (0.78%) |
| 15 | "Flow" | 7045 (3.3%) | "Aqueous-solution" | 4805 (2.9%) | "Oxidation" | 1108 (0.78%) |
| 16 | "Reactor" | 6977 (3.3%) | "Mechanism" | 4324 (2.6%) | "Nanoparticles" | 1102 (0.77%) |
| 17 | "Particle" | 6706 (3.1%) | "Particle" | 4078 (2.4%) | "Activated carbon" | 1081 (0.76%) |
| 18 | "Production" | 6562 (3.1%) | "Reactor" | 4049 (2.4%) | "Diffusion" | 1080 (0.76%) |
| 19 | "Method" | 6294 (3.0%) | "Simulation" | 4027 (2.4%) | "Methane" / "CH4" | 1038 (0.73%) |
| 20 | "Catalytic" | 6242 (2.9%) | "Combustion" | 3979 (2.4%) | "Hydrogen" / "H2" | 1018 (0.71%) |
| 21 | "Application" | 6089 (2.9%) | "Surface" | 3784 (2.2%) | "Separation" | 1003 (0.70%) |
| 22 | "Performance" | 5957 (2.8%) | "Design" | 3721 (2.2%) | "Combustion" | 968 (0.68%) |
| 23 | "Surface" | 5952 (2.8%) | "Acid" / "Acidity" | 3638 (2.2%) | "Mixing" | 963 (0.68%) |
| 24 | "Liquid" | 5947 (2.8%) | "Catalysts" | 3388 (2.0%) | "Fouling" | 955 (0.67%) |
| 25 | "Oxidation" | 5922 (2.8%) | "Waste-water" | 3276 (1.9%) | "Biomass" | 948 (0.66%) |
| 26 | "Characterization" | 5839 (2.7%) | "Degradation" | 3113 (1.9%) | "Biodiesel" / "biodiesel fuel" | 915 (0.64%) |
| 27 | "Oil" | 5749 (2.7%) | "Mass-transfer" | 2927 (1.7%) | "Ethanol" / "bioethanol" | 907 (0.64%) |
| 28 | "Control*" | 5635 (2.6%) | "Optimization" | 2913 (1.7%) | "Heat transfer" | 907 (0.64%) |
| 29 | "Reaction" | 5599 (2.6%) | "Methane" / "CH4" | 2872 (1.7%) | "Extraction" | 847 (0.59%) |
| 30 | "Adsorption" | 5749 (2.6%) | "Transport" | 2728 (1.6%) | "Desalination" | 840 (0.59%) |

Next words are “process” (ranked 3$^{th}$), “systems” (ranked 4$^{th}$) and “properties” (ranked 5$^{th}$), which are very general words to be considered for the identification of research areas and trends. They were followed by “model” and “modeling/modelling”, ranked 7$^{th}$ and 8$^{th}$, respectively. In fact, if “model” and “modeling/modelling” would be counted together, they will be the first word ranked. This means mathematical modelling is present in almost all the research fields of Chemical Engineering, as demonstrated by the extended use of process design and simulation software in the daily practice. The importance of “model” decreased slightly during the analyzed period, from 4.4% to 4.1% total publications, while the importance of “modeling/modelling” increased in a similar extent, i.e. from 4.0% to 4.4% total publications, meaning the use of mathematical modelling is already a mature area in Chemical Engineering.

Other frequent words were “water” and “carbon”, ranked 10$^{th}$ and 11$^{th}$, respectively. Water treatment is becoming an area of increased importance in Chemical Engineering. In fact, its share in total words in the articles titles increased largely during the analyzed period, from 2.2% in 2000 to 4.6% in 2011. On the other hand, the importance of the word “carbon” can be due to the use of activated carbon as sorbent, i.e. for wastewater treatments (also reflected by the importance of word “adsorption”, ranked 30$^{th}$), or by the increased interest in carbon dioxide mitigation technologies against climate change. The increased importance of “carbon” was reflected by the increased share in article titles from 2.8% in 2000 to 4.9% in 2011.

Other frequent words used in the article titles were “kinetics” (ranked 14$^{th}$), “reactor” (ranked 16$^{th}$), “catalytic” (ranked 20$^{th}$) and “reaction” (ranked 29$^{th}$), all terms related to chemical reaction engineering, one of the most important areas of Chemical Engineering, which was also demonstrated by the 2$^{nd}$ rank of the word “catalyst”. The importance of “reactor” or “catalytic” is similar along the analyzed period but the importance of “kinetics” have increased largely, from around 2.9% in 2000 to 3.7% in 2011.

“Surface” is ranked 23$^{th}$, indicating the surface phenomena and surface properties are also important for chemical engineers. Word “oxidation” (ranked 26$^{th}$) indicates this is one of the chemical reactions of a larger interest. It involves catalytic oxidation, partial oxidation and the advanced oxidation processes for wastewater treatment. The importance of both words, "surface" and "oxidation", has only varied slightly along the analyzed period.

Keyword Plus®. There were a total of 168,630 articles with keyword plus® available in their bibliographic information, i.e. 78.5% of total publications in 2000-2011, this coverage increasing from 57.1% in 2000 to 90.1% in 2011. The average of keyword plus® per article was 6.3 in the period 2000-2011, this number increasing from 4.9 in 2000 to 7.3 in 2011.

In this case, "system" is ranked 1st and "model" is ranked 2nd (see Table 1). In keyword plus®, the words "modeling" or "modelling" are almost negligible, despite these words were frequently used by the authors in the articles titles. This means the word "model" probably include both "model" and "modeling/modelling".

Some of the most relevant topics observed in the analysis of title words appeared again in the analysis of keyword plus® but they were ranked in first positions, i.e. "oxidation" (ranked 3rd), "water" (ranked 4th) and "kinetics" (ranked 5th). "Adsorption" is ranked 6th. Although it was ranked 30th in the 30 most frequently words used in article titles, it is well known that this is one of the most important unit operations in Chemical Engineering. The analysis of keyword plus® allowed distinguishing the relative importance between activated carbon and carbon dioxide as research areas. The importance of carbon dioxide (ranked 11th, 5,292 publications, 3.1% total publications) was clearly higher than that of activated carbon (not among the 30 most frequently used keyword plus®, 2,134 publications, 1.27% total publications), indicating the importance of climate change research for Chemical Engineering. Again, words as "reactor" (ranked 18th) and "catalysts" (ranked 24th), together with the previously mentioned "kinetics" (ranked 5th), corroborated the importance of chemical reaction engineering in Chemical Engineering.

Furthermore, some new keywords appeared in the list of the 30 most frequently keyword plus®: "simulation" (ranked 19th), "combustion" (ranked 20th), "waste-water" (ranked 25th), "mass-transfer" (ranked 27th), "methane" + $CH_4$" (ranked 29th) and "transport" (ranked 30th).

Author keywords. There were 142,607 articles with available author keywords (66.9% total publications), the share of these articles increasing from 44.8% in 2000 to 76.2% in 2011. The number of keywords used for the authors was similar along the analyzed period (average 4.9). The number of author keywords is usually limited by the journals and thus it is lower than the number keyword plus®: 4.9 vs. 6.3.

"Modeling/modelling" (ranked 1st), "simulation" (ranked 5th), "CFD/computational fluid dynamics" (ranked 6th) and "mathematical modeling/modelling" (ranked 12th) were among the most frequently used keywords, indicating the importance of modelling and computational methods to solve problems in Chemical Engineering (see Table 1). Their importance was very similar along the analyzed period, indicating this is an already consolidated area, which was in agreement with the results obtained by the analysis of words in article titles or keyword plus®. On the contrary, the use of computational fluid dynamics (CFD) has increased largely, from 0.84% in 2000 to 1.5% in 2011, and in a lower extent, "mathematical modeling", from 0.84% in 2000 to 1.1-1.2% in 2010-11.

"Adsorption" was the second most used keyword and "activated carbon", the most typical sorbent in adsorption, was ranked 17th. Again, the importance of carbon dioxide (ranked 7th) is higher than "activated carbon", as previously suggested by the analysis of keyword plus®. The importance of chemical reaction engineering was also demonstrated by the importance of the keyword "kinetics" (ranked 3rd), however, no words such "catalyst" or "reactor" were among the top 30 most used by the authors as occurs when words in article titles or keyword plus® were analyzed, probably because these are very general terms to be used as author keywords.

Contrary to article title words or keyword plus®, different types of membrane treatments appeared among the 30 most frequently used author keywords such as "ultrafiltration" (ranked 10th), "nanofiltration" (ranked 13th) and "reverse osmosis" (ranked 14th). All of them had a very similar importance, also similar to the generic word "membrane" (ranked 11th). The analysis of author keywords in this case highlights the present importance of membranes in Chemical Engineering. The keywords "fouling" (ranked 24th) and "desalination" (ranked 30th) also corroborates the increased importance of wastewater treatment and drinking water production, especially by membrane treatments. The share of these words in publications was almost the same during the analyzed period.

"Mass-transfer" became one of the most important keywords used by the authors (ranked 7th), while was ranked only 27th in keyword plus®. "Mass-transfer" together with "diffusion" (ranked 18th), "mixing" (ranked 23rd) and "heat transfer" (ranked 28th) shown the importance of transport phenomena in Chemical Engineering. Again, it was confirmed that "oxidation" and "combustion" are the two types of chemical reactions with a higher interest in Chemical Engineering: “oxidation” ranked 15$^{th}$ and combustion ranked 22$^{nd}$. On the other side, it has been observed “photocalysis” is becoming an important area in Chemical Engineering (ranked 9$^{th}$), which was not detected either by the analysis of title words the keyword plus®. In fact, the importance of photocatalytic processes have increased largely during the analyzed period, from 0.42% in 2000 to 1.6% total publications in 2011.

Hydrogen was also an important research topic (ranked 20$^{th}$), not detected either in title words or keyword plus®. The importance of hydrogen as a new energy vector and the use of fuel cells for energy production is becoming a research field of a great interest for chemical engineers. The importance of “hydrogen” as keyword have increased largely from 0.24% in 2000 to 0.77% total publications in 2011.

Two new relevant areas, with very rapid increases, were also detected by the analysis of author keywords: “nanoparticles” (ranked 16$^{th}$) and “biodiesel” (ranked 26$^{th}$). The keyword “nanoparticles”, for example, appeared only in 12 articles in 2000 but in 207 articles in 2011, increasing its presence from 0.19% in 2000 to 1.1% total publications in 2011. Similarly, “biodiesel” was used as author keyword in 286 articles in 2011 (1.5% of total articles), while the first papers appeared in 2001.

Finally, “biomass” (ranked 25$^{th}$) and “ethanol”/”bio-ethanol” (ranked 27$^{th}$) were frequently used as author keywords. These keywords were not among the most frequently used neither in article title or keyword plus®. However, the increased use of biomass as a new raw material and energy source, and the production of ethanol by fermentation, are actually some of the most interesting areas for Chemical Engineering researchers.

### 3.2. Analysis of the 1,000 most cited papers

Words in articles title. All articles have title (100% of total publications). The total number of words analyzed was 11,450, giving an average number of 11.4 words per article. Again prepositions, conjunctions, etc., were removed before the analysis. They represented 26.7% of total raw words. A total of 2,622 different words were obtained. A large percentage of words appeared only once (1577 words, 60.1% total words), 376 words appeared twice (14.3%), 174 words appeared tree times (6.64%) and 496 words appeared four or more times (18.9%).

Table 2 shows the 30 most frequently single words used in article titles. As observed, still some words in the first positions could be considered as low relevant for determining research areas such as "review" (3rd ranked), "process" (9th ranked) or "applications" (10th ranked). The word "review", however, highlights the importance of reviews among the most cited papers. It is well known that review papers tend to be cited more frequently than other types (Aksnes, 2003; Moed, 2010; Vanclay, 2013; Miranda and Garcia-Carpintero, 2018), however, the overrepresentation of reviews in the most cited papers in Chemical Engineering is huge. As demonstrated Part 2 of this study, a 65% of the 1,000 most cited papers were reviews while their share in publications was only 1.5% (Miranda and Garcia-Carpintero, 2020b).

"Catalysts" is the most used word in article titles of the most cited papers of the area. This word, together with "catalytic" (ranked 6th), "photocatalytic" (ranked 8th), or "activity" (ranked 21st), indicates the importance of chemical reaction engineering and catalytic reactions in Chemical Engineering. "Membranes" is the second most frequent word, which reflects its importance in the area, for both wastewater and drinking water production (desalination) and fuel cells. Related to fuel cells, it is important to consider that the word "fuel" is ranked 4th, however, the word cannot be specifically related to fuel cells but it can be related to other types of fuels (coal, petroleum, etc.). However, the frequency of "cell" (ranked 28th) and "methanol" (ranked 30th) could justify the importance of fuel cells area. The joint analysis with keyword plus® or author keyword will be necessary to determine the exact importance of fuel cells.

Again "oxidation" (ranked 5th) and "combustion" (ranked 13th) appeared as some of the most important types of chemical reactions, however, photocatalytic reactions outstand as a relatively new area, already 8th ranked. Besides, "photocatalytic" is related to some other words such as "$TiO_2$" / "titania", which is the most popular photocatalyst, which was found also among the top 30 most frequently words (ranked 20th). The word "carbon" appeared ranked 7th. As commented earlier this could be related to carbon dioxide or activated carbon, even to some new emerging materials such as carbon nanotubes.

**Table 2.** Top 30 words in article titles, keyword plus® and author keywords of the 1,000 most cited papers.

| Rank | Words in title | Publications | Keyword plus® | Publications | Author keywords | Publications |
|---|---|---|---|---|---|---|
| 1 | "Catalysts" | 153 (15.3%) | "Kinetics" | 63 (6.5%) | "Adsorption" | 43 (5.4%) |
| 2 | "Membranes" | 95 (9.5%) | "Adsorption" | 63 (6.5%) | "Titanium dioxide"/ "$TiO_2$" / "Titania" | 42 (5.3%) |
| 3 | "Review" | 87 (8.7%) | "Water" | 58 (6.0%) | "Photocatalysis" | 36 (4.5%) |
| 4 | "Fuel" | 79 (7.9%) | "$TiO_2$" / "Titanium dioxide" | 57 (5.9%) | "Kinetics" | 34 (4.3%) |
| 5 | "Oxidation" | 69 (6.9%) | "Aqueous-solutions" | 51 (5.3%) | "Fuel cell" | 32 (4.0%) |
| 6 | "Catalytic" | 60 (6.0%) | "Oxidation" | 50 (5.2%) | "Biodiesel" | 23 (2.9%) |
| 7 | "Carbon" | 54 (5.4%) | "Carbon-dioxide" / "$CO_2$" | 48 (5.0%) | "Gold" / "Au" | 20 (2.5%) |
| 8 | "Photocatalytic" | 52 (5.2%) | "Removal" | 44 (4.5%) | "Biosorption" | 16 (2.0%) |
| 9 | "Process" | 51 (5.1%) | "Carbon-monoxide" / "CO" | 43 (4.4%) | "Modeling" / "modelling" | 16 (2.0%) |
| 10 | "Applications" | 50 (5.0%) | "Waste-water" | 42 (4.3%) | "Catalysis" | 15 (1.9%) |
| 11 | "Adsorption" | 48 (4.8%) | "Degradation" | 39 (4.0%) | "Nanoparticles" | 15 (1.9%) |
| 12 | "Removal" | 42 (4.2%) | "Systems" | 37 (3.8%) | "Photocatalyst" | 15 (1.9%) |
| 13 | "Combustion" | 41 (4.1%) | "Surface" | 35 (3.6%) | "Carbon dioxide" / "CO2" | 15 (1.9%) |
| 14 | "Aqueous" | 40 (4.0%) | "Temperature" | 34 (3.5%) | "Hydrogen production" | 14 (1.8%) |
| 15 | "Properties" | 40 (4.0%) | "Mechanism" | 33 (3.4%) | "Catalyst" | 14 (1.8%) |
| 16 | "Ionic" | 39 (3.9%) | "Performance" | 31 (3.2%) | "Isotherms" | 14 (1.8%) |
| 17 | "Water" | 39 (3.9%) | "Sorption" | 30 (3.1%) | "CO" / "carbon monoxide oxidation" | 14 (1.8%) |
| 18 | "Characterization" | 38 (3.8%) | "Model" | 30 (3.1%) | "Ionic liquids" | 14 (1.8%) |
| 19 | "Production" | 37 (3.7%) | "Fuel-cell" or "fuel-cell applications" | 30 (3.1%) | "Wastewater Treatment" | 13 (1.6%) |
| 20 | "$TiO_2$" / "titania" | 37 (3.7%) | "Activated carbon" | 29 (3.0%) | "Ceria" | 12 (1.5%) |
| 21 | "Reaction" | 36 (3.6%) | "Hydrogen" | 28 (2.9%) | "Proton conductivity" | 12 (1.5%) |
| 22 | "Study" | 36 (3.6%) | "Catalyst" | 28 (2.9%) | "Steam reforming" | 12 (1.5%) |
| 23 | "Synthesis" | 36 (3.6%) | "Reduction" | 26 (2.7%) | "Membrane" | 12 (1.5%) |
| 24 | "Activity" | 35 (3.5%) | "Methane" / "$CH_4$" | 26 (2.7%) | "Oxidation" | 11 (1.4%) |
| 25 | "Hydrogen" | 35 (3.5%) | "Mixture" | 26 (2.7%) | "Ethanol" / "bioetanol" | 11 (1.4%) |
| 26 | "Biodiesel" | 34 (3.4%) | "Acid" | 26 (2.7%) | "Palladium" / "Pd" | 11 (1.4%) |
| 27 | "Acid" | 34 (3.4%) | "Particle" | 26 (2.7%) | "Hydrogen" | 11 (1.4%) |
| 28 | "Cell" | 32 (3.2%) | "Separation" | 23 (2.4%) | "Platinum" / "Pt" | 11 (1.4%) |
| 29 | "Liquids" | 32 (3.2%) | "Decomposition" | 22 (2.3%) | "Pyrolysis" | 10 (1.3%) |
| 30 | "Methanol" | 31 (3.1%) | "Behavior" | 21 (2.2%) | "Fouling" | 10 (1.3%) |

Notes: singular and plurals were considered together; percentages are based on publications with available information, i.e. 1,000 publications for title words, 969 for keyword plus® and 793 for author keywords.

Next significant word is "adsorption" (ranked 11th). As previously seen, adsorption is one of the most frequently used unit operations. It is also interesting the word "ionic" (ranked 16th), which is a very general term, the same as "liquids" (ranked 29th), however these two words together could be related to ionic liquids, which is one present hot topic in Chemical Engineering. Again the analysis of keyword plus® or author keywords will determine how important is this area in Chemical Engineering. Finally, "biodiesel" appeared 26th ranked. This word was not detected by the analysis of article title words or keyword plus® in total scientific production.

Keyword plus®. There were 969 articles with keyword plus® (96.9% of total publications). The total number of words analyzed was 7,811, giving an average number of 8.1 keyword plus® per article. The number of different words were 4,270, most of them appeared only once (3185 words, 74.6%), 542 appeared two times (12.7%) and 209 appeared three times (4.9%). Only 334 words (7.8%) appeared more than three times. In this case, there was a greater amount of generic keywords of little interest to identify research areas such as "water", "aqueous-solutions", "removal", "degradation", "systems", "performance", "behavior", etc.

More frequent keyword plus® were "kinetics" and "adsorption", with the same number of counts (see Table 2). "Kinetics" together with "catalyst" (ranked 22nd), corroborated the importance of chemical reaction engineering. Besides, "adsorption" together with "sorption" (ranked 17th), "activated carbon" (ranked 20th) confirmed the importance of adsorption as unit operation and activated carbon as the most frequently used sorbent.

The word "waters" was ranked 3rd. This term is very general but if we consider the importance of other keywords such as "waste-water" (10th ranked), possibly also "removal" (ranked 8th) or "degradation" (ranked 11th), there is a clear importance of wastewater treatment in Chemical Engineering. "Titanium dioxide" is the fourth ranked keyword, which is the most commonly used photocatalyst in advanced oxidation processes. Again the importance of photocatalysis, and titanium dioxide as photocatalyst, is demonstrated. Although "combustion" is not among the top 30 keyword plus®, "oxidation" again appears as one of the most important chemical reactions, in this case, ranked 6th. "Carbon dioxide" is ranked now in 7th position, which is consistent with previous findings, however, now "carbon monoxide" appeared for the first time (9th ranked). "Surface" appeared in 13rd position, "model" appeared in 18th position, "hydrogen" appeared in 21st position and "methane" in 24th.

Other important finding is the 19th position of "fuel cells" or "fuel cells applications", which appeared for the very first time, confirming the origin of the high frequencies of "fuel" and "cells" in the analysis of the words of the article titles.

Author keywords. There were 793 articles with author keywords (79.3% of total publications). The total number of words analyzed was 4,476, giving an average number of 5.6 words per article. The number of different words were 3,110. Most of them appeared only once (2540 words, 81.7%), 326 appeared two times (10.5%) and 91 appeared three times (2.9%). Only 153 words (4.9%) appeared more than three times. As shown in Table 2, generic terms such as "review", "process", "applications", did not appeared. Furthermore, it was not necessary to remove prepositions, conjunctions, etc. As occurred when total scientific production was analyzed, the number of generic terms was significantly lower than keyword plus®.

More frequent keyword is "adsorption". This word, together with "biosorption" (8th ranked), "isotherms" (15th ranked) and "activated carbon" (31th ranked, not shown), indicates the large importance of unit operations, and specifically adsorption, for a variety of areas of Chemical Engineering. Furthermore, the use of photocatalytic processes outstands in author keywords with "titanium dioxide" (ranked 2nd), "photocatalysis" (ranked 3rd) but also with "photocatalyst" (ranked 12th).

As demonstrated in previous analyses, the importance of chemical reaction engineering is very high as demonstrated by the presence of keywords "kinetics" (ranked 4th), "catalysis" (10th ranked), and all the keywords related to advanced oxidation processes and photocatalysis. Fuel-cells and biodiesel appeared for the very first time in the top positions, in position 5th and 6th, respectively. Furthermore, "Proton conductivity" (ranked 20th), related to the membranes used for fuel cells applications, appeared for the first time in this study. "Nanoparticles" was ranked 11th. Some of these nanoparticles are noble metals such as "gold" (ranked 7th), "palladium" (ranked 25th), "platinum" (ranked 27th) or "ceria" (ranked 19th).

"Modeling" was 9th ranked. "Hydrogen production" is 13th ranked, while "hydrogen" was 26th ranked, demonstrating the importance of hydrogen in the area. "Carbon monoxide oxidation" was 16th ranked. Again, "ionic liquids" appeared in the top 30. In this case, ranked 17th. "Wastewater treatment" was 18th ranked. The keyword "membranes" is in the list of top 30 (22nd ranked), however, in a lower position than in other previous analysis. The presence of "fouling", 29th ranked, usually employed for fouling studies of membranes, also highlight the importance of membranes, especially for wastewater treatments or drinking water production. Finally, "Oxidation" is 23rd ranked and "Ethanol/Bioethanol" is 24th ranked.

### 3.3. Research trends and hot topics

In the previous sections, the combined analysis of the most frequently words in article title, keyword plus® and author keywords gave an approximate idea of what are the most important research topics in Chemical Engineering, i.e. chemical reaction engineering, unit operations (e.g. adsorption), membrane treatments, advanced oxidation processes, carbon dioxide and climate change, etc.

Furthermore, 10 areas outstand from the previous analysis, characterized by a low number of publications but a very rapid increase with time. These areas can be considered as the main hot topics of the area in the period 2000-2011: hydrogen, wastewater, carbon dioxide, photocatalysis, nanoparticles, biodiesel, ionic liquids, membranes, fuel cells and biomass. All they have in common a very rapid development in the last decade, as shown in Figure 1.

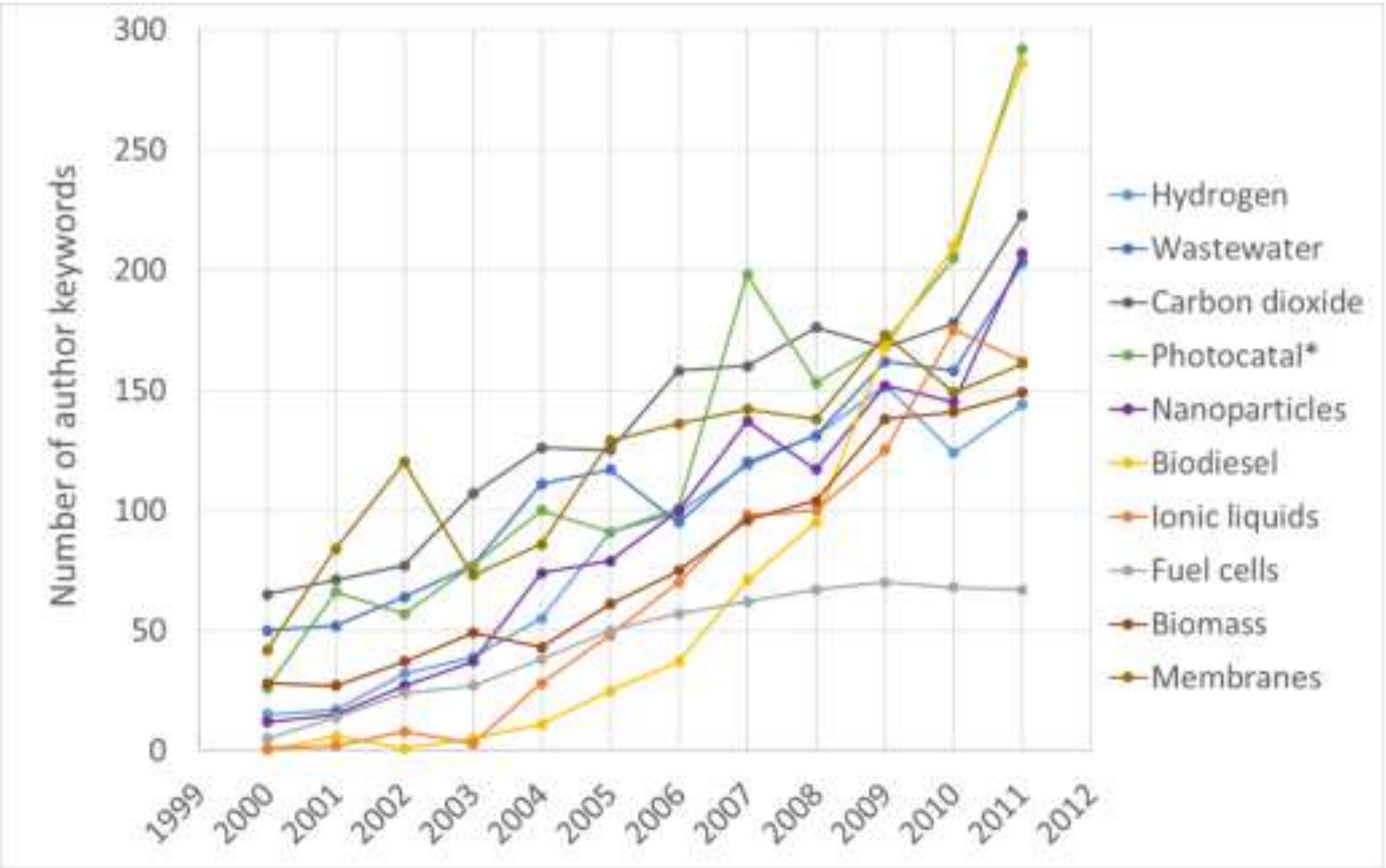

**Figure 1.** Evolution of the number of publications identified as the most innovative research areas and hot topics in Chemical Engineering in 2000-2011.

Table 3 summarizes the coverage of these areas by the different lists of the top 30 most frequent keywords obtained in this study. As it can be seen, the analysis of the 1,000 most cited papers demonstrated to be the most efficient approach to detect research areas which were not previously identified by the analysis of the total publications. As occurs with total publications, author keywords were the most useful for identifying research hot topics.

### 3.4. Comparison of results and recommendations

<u>Total scientific production or highly cited papers?</u> First, there is a large difference in the number of publications to be analyzed. If total scientific production is analyzed, a large number of publications needs to be considered, depending on the area and the time period considered. In the present case, the total number of documents (articles plus reviews only) was 213,264, while the most cited publications analyzed were only 1,000. Even considering the 1% most cited publications, the difference in the number of publications would be huge (2,133 publications). The larger number of publications to be analyzed implies not only a larger time-consuming analysis of words, but also a time-consuming activity of downloading the bibliographic information from Web of Science database. At present, a maximum of 500 records can be downloaded at the same time from this database.

The computing time necessary to analyze the thematic areas in total scientific production year by year is also considerably higher than for most cited papers and powerful computers are necessary to avoid the computer freezes up during the calculations. Even more important, although some degree of automatization in the analysis of the frequency of appearance of words is possible through Microsoft Excel® tools, there is mandatory a further manual analysis of the results. This analysis can be very complex and time-consuming, i.e. considering singular, plurals, abbreviations, the use of chemical formulas, etc., but necessary. In fact, some words, which were not initially in the raw top 30 most frequent words, appeared in first positions of the list after refining.

For detecting research areas trends or hot spots the most cited papers can be used instead, reducing largely the necessary time for the analyses. As also recognized by Ho *et al.* (2010) when determining the hotspots of Japanese lung cancer research, hot spots can be efficiently reflected by some highly cited papers. The analysis of total scientific production usually highlights only mature and large research areas.

**Table 3**. Comparison of the coverage of the main hot topics in Chemical Engineering by different methodological approaches.

| | Total scientific production | | | 1,000 most cited papers | | |
|---|---|---|---|---|---|---|
| | **Top words in article title** | **Top keyword plus®** | **Top author keyword** | **Words in article title** | **Keyword plus®** | **Author keyword** |
| **Hydrogen** | - | - | "hydrogen" (20th) | "hydrogen" (25th) | "hydrogen" (21th) | "hydrogen production" (13th), "hydrogen" (26th) |
| **Wastewater** | "water" (12th) | "water" (4th)<br>"waste-water" (25th) | "wastewater" (30th) | water" (17th) | "wastewater" (10th) | "wastewater treatment" (18th) |
| **Carbon dioxide** | "carbon" (13th) | "carbon dioxide" (11th) | "carbon dioxide" (8th) | "carbon" (7th) | "carbon dioxide" (7th) | "carbon dioxide" (12th) |
| **Photocatalysis/ photocatalyst** | - | - | "photocatalysis"/ "photocatalyst" (9th) | "photocatalytic" (8th), "titanium dioxide" (20th) | "titanium dioxide" (4th) | "titanium dioxide" (2nd)<br>"photocatalysis" (3rd)<br>"photocatalyst" (12th) |
| **Nanoparticles** | - | - | "nanoparticles" (16th) | - | - | "gold" (7th)<br>"nanoparticles" (11th) |
| **Biodiesel** | - | - | "nanoparticles" (26th) | "nanoparticles" (26th) | - | "nanoparticles" (6th) |
| **Ionic liquids** | - | - | - | "ionic" (16th)<br>"liquids" (29th) | - | "ionic liquids" (17th) |
| **Membranes** | - | - | "ultrafiltration" (10th)<br>"nanofiltration" (13th)<br>"reverse osmosis" (14th),<br>"fouling" (24th) | "membranes" (2nd) | - | "membrane" (22th)<br>"fouling" (28th), "proton conductivity" (20th) |
| **Fuel cells** | - | - | - | "membranes" (2nd)<br>"fuel" (4th)<br>"cell" (28th)<br>"methanol" (30th) | "fuel-cell" or "fuel-cell applications" (19th) | "fuel cells" (5th)<br>"proton conductivity" (20th) |
| **Biomass** | - | - | "biomass" (25th) | - | - | - |

Words in article title, keyword plus® or author keywords? One drawback of using words in article title is the presence of a high number of non-significant words consisting in prepositions, conjunctions, etc., which must be removed from the analysis. When analyzing the author keywords or keyword plus®, this drawback does not exist, minimizing the processing of not useful data. This means that, although the number of words processed in article titles is higher, the number of different keywords is higher for keyword plus® and author keywords.

Other important drawback when using the analysis of words in articles title is that only single words are considered. In some research areas, two or more words are necessary to define exactly which is the research area. These words can be very common theirselves, however, the combination is unique. Many examples have been described along this study such as “ionic liquids”, “fuel cells”, “activated carbon”, “carbon dioxide”, etc. This is not the case when analyzing keyword plus® or author keyword. Contrary to words in article title, in author keywords, the intact words that the authors want to transmit to the readers are preserved (Li et al., 2009).

One advantage of the analysis of words in article titles is that 100% of the documents have a complete title, while the coverage of keyword plus® and author keyword were 78.5% and 66.9% for total publications, respectively. However, the missing data for keyword plus® and author keyword have decreased significantly from 2000 to 2011 and it is expected to be a minor problem in the next years, for example a 90.1% of the articles published in 2011 already had keyword plus® and a 76.2% had author keywords.

One common drawback for all the keywords is the high percentage of keywords appearing only once. In the case of total scientific production, 79.8% of words in titles appeared three or less times, and this percentage increased to 86.9% for keyword plus® and 92.6% for author keywords. This fact has been also previously observed by other authors in different research areas. For example, Tian *et al.* (2008), obtained analyzing the author keywords for Global Information System (GIS) research that a 78% of the author keywords appeared only once, 11% were used twice and 3.9% appeared three times. Fu *et al.* (2010), in the area of solid waste research, found that a 61% of author keywords were used only once, a 14% were used twice, 6.6% were used three times, and only 18% keywords were used more than three times. In other studies, the results are practically the same: a 73% of author keywords were used just once and 11% were used twice in the area of obstructive sleep apnea (Huang, 2009), a 74% of author keywords which appeared once and 11.3% twice in the area of nanotribology (Elango *et al.*, 2013), a 74% of author keywords appeared once and 11% appearing twice in the area of aerosol research (Xie *et al.*, 2008) and 70.8% of author keywords appeared once and 11.7% appeared twice in the stem cell research (Li *et al.*, 2009).

For explaining the large number of keywords used only once, most of the authors (Tian *et al.*, 2008; Xie *et al.*, 2008; Li *et al.*, 2009; Fu *et al.*, 2010; Elango *et al.*, 2013) accept the explanation first given by Chuang *et al.* (2007), i.e. this is an indication of the lack of continuity in research and a wide disparity in research focus. However, these results can be explained by the fact that there are many ways to formulate different research areas and many grammatical variations which are very difficult to be considered when millions of words are analyzed. There are many ways to express the same, especially when chemical formulas, abbreviations, etc. are common in practice. In addition, if a careful pre-treatment of the raw data is not carried out, some keywords can be not counted together, i.e. the presence of a space before the keyword, all capital letters vs. all small letters, first letter of the keyword with capital letters or not, the use of "-" to form combined words, keywords with one, two or three words will not be counted together even they are related to the same research area, etc. Difficulties in this type of analysis have been previously indicated, for example, by Xie *et al.* (2008), "the comparisons of the keywords frequency would somewhat produce inexact or spurious conclusions, i.e. the term PM 10 had three expressed forms: PM-10, PM(10) and PM 10, and even "inhalable particles" were with the same sense as author keywords."

The number of author keywords is lower than the number of keyword plus®. If 10 keyword plus® are allowed per publication, the author keywords are usually limited by the journals to 4-6. Although this could be a drawback, the truth is that author keywords are less general and more specific and meaningful than keyword plus®.

Authors analyzing together words in article title, keyword plus® and author keyword found that the ranking of the most frequently used keywords fluctuated only slightly (Xie *et al.*, 2008), i.e. 16 of the top 20 author keyword title word or abstract word, appeared in the top 20 keyword plus® (Zhang *et al.*, 2010a) or 14 of the top 30 author keywords and words in title were also identified by keyword plus® (Ho *et al.*, 2010). J. Zhang *et al.* (2016), analyzing the patient adherence research, obtained that keyword plus® were as effective than author keywords in terms of investigating the knowledge structure of scientific fields, but it was less comprehensive in representing an article's content. According to their developers, keyword plus® is usually more concerned about novel research directions than the mature direction in the field (Garfield & Sher, 1993), however, in our study, the most valuable information was given by author keywords, keyword plus® still being more relevant than words in article title.

In this study, the presence of irrelevant words in the top frequent keywords was more important when words in article title or keyword plus® were analyzed than for author keywords. This is opposite to the results obtained by Fu *et al.* (2010), in the area of solid waste research, and Huang (2009) in the area of obstructive sleep apnea. They found keyword plus® identified additional terms which describe the articles' content with greater variety which were not apparent among other keywords. As a summary, Table 4 compares the characteristics and results of the different methodological approaches carried out in this study.

**Table 4**. Comparison between the analysis of words in article titles, keyword plus® and author keywords.

| | Words in article title | Keyword plus® | Author keyword |
|---|---|---|---|
| Is possible keywords of two or more words? | No | Yes | Yes |
| Analysis by publication years | Yes | Yes | Yes |
| Covering of the total publications | 100% | 78.5% (57.1% in 2000 → 90.1% in 2011) | 66.9% (44.8% in 2000 → 76.2% in 2011) |
| Number of words to be analyzed | 2,558,032 | 1,980,549 | 699,217 |
| Different keywords | 287,735 (11.2%) | 315,122 (15.9%) | 371,666 (53.1%) |
| Average words per publication | 12.0 (10.9 → 12.8) | 6.3 (4.9 → 7.3) | 4.9 (4.7 → 5.0) |
| Frequency of words | 61.5% appeared once; 12.6% two times; 5.7% three times | 67.5% appeared once; 13.8% two times; 5.6% three times | 78.4% appeared once; 10.4% two times; 3.8% three times |
| Efficiency in detecting hot topics | ↑↑ | ↑↑ | ↑↑↑ |
| Complexity of the analysis | Intermediate (depending on the number of documents considered) | Intermediate (depending on the number of documents considered) | Intermediate (depending on the number of documents considered) |

Note: The numerical values included in the table are calculated for total scientific production, however, the trends are approximately the same for the 1,000 most cited papers.

## 4. CONCLUSIONS

The analysis of words in article title, keyword plus® and author keywords were carried out both in total scientific production and in the most cited papers. While the analysis of total scientific production gives an idea of the most important (and mature) areas of the field, the highly cited papers are more efficient for detecting the hot topics of the area. Words like "hydrogen", "nanoparticles", "biodiesel", "biomass", "ethanol"/"bioethanol" (and different membrane issues) were not detected among the top 30 most frequent words neither in article titles or keyword plus® and only in some author keywords. The analysis of the highly cited papers is also useful to obtain the most important and mature research areas while reducing largely the time required for the analysis.

Words in article title, keyword plus® and author keywords are complementary, however, author keywords are suggested as the source of most useful data. Compared to the other approaches, the author words are more significant for identifying research areas and trends and the total number of words to be analyzed is lower. The main drawback of using words in article title is that they are single words, and many research areas are better defined by a combination of at least two words. On the other hand, the keyword plus® were less efficient than the other words analysis for identifying research trends and especially hot topics, probably because they are obtained as frequently occurring words in the titles of an article's cited references, with implies a certain delay.

Although automatic tools from different software such as Microsoft Excel® can be helpful for automatic counting of the words, a further careful manual processing is always necessary. Finally, the authors recommend a revision of the bibliometric analysis of words by experts in the area to avoid inaccurate results and to obtain the most valuable information.